\documentclass[iop,apj,tighten]{emulateapj}
\usepackage{apjfonts} 
\usepackage{amsmath,amstext}
\usepackage[breaklinks,colorlinks,citecolor=black,linkcolor=black,urlcolor=black]{hyperref} 
\usepackage[all]{hypcap} 

\submitted{Received 2016 August 31; revised 2016 September 23; accepted 2016 October 3}
\journalinfo{The Astrophysical Journal Letters}

\newcommand{\Ni}{\ensuremath{^{56}\mathrm{Ni}}}

\newcommand{\Mej}{\ensuremath{M_\mathrm{ej}}}
\newcommand{\Mni}{\ensuremath{M_\mathrm{\Ni}}}
\newcommand{\Eej}{\ensuremath{E_\mathrm{ej}}}

\newcommand{\Msun}{\ensuremath{M_\odot}}
\newcommand{\vw}{\ensuremath{v_w}}

\newcommand{\kmps}{\ensuremath{\mathrm{km~s^{-1}}}}

\newcommand{\ej}{\ensuremath{\mathrm{ej}}}
\newcommand{\Mloss}{\ensuremath{\dot{M}_\mathrm{loss}}}
\newcommand{\Macc}{\ensuremath{\dot{M}_\mathrm{acc}}}
\newcommand{\rhocsm}{\ensuremath{\rho_\mathrm{CSM}}}
\newcommand{\sh}{\ensuremath{\mathrm{sh}}}
\newcommand{\rsh}{\ensuremath{r_\mathrm{sh}}}
\newcommand{\vsh}{\ensuremath{v_\mathrm{sh}}}

\shorttitle{Radio transients from AIC}
\shortauthors{Moriya}

\begin{document}

\title{Radio transients from accretion-induced collapse of white dwarfs}
\author{Takashi J. Moriya}
\affil{Division of Theoretical Astronomy, National Astronomical Observatory of Japan, National Institues of Natural Sciences, \\ 2-21-1 Osawa, Mitaka, Tokyo 181-8588, Japan \\ takashi.moriya@nao.ac.jp}

\begin{abstract}
We investigate observational properties of accretion-induced collapse (AIC) of white dwarfs in radio frequencies. If AIC is triggered by accretion from a companion star, a dense circumstellar medium can be formed around the progenitor system. Then, the ejecta from AIC collide to the dense circumstellar medium, making a strong shock. The strong shock can produce synchrotron emission which can be observed in radio frequencies. Even if AIC occurs as a result of white dwarf mergers, we argue that AIC may cause fast radio bursts if a certain condition is satisfied. If AIC forms neutron stars which are so massive that rotation is required to support themselves (i.e., supramassive neutron stars), the supramassive neutron stars may immediately lose their rotational energy by the r-mode instability and collapse to black holes. If the collapsing supramassive neutron stars are strongly magnetized, they may emit fast radio bursts as previously suggested. The AIC radio transients from the single-degenerate systems may be detected in the future radio transient surveys like Very Large Array Sky Survey or the Square Kilometer Array transient survey. Because AIC is suggested to be a gravitational wave source, gravitational waves from AIC may be accompanied by radio-bright transients which can be used to confirm the AIC origin of observed gravitational waves.
\end{abstract}

\maketitle

\section{Introduction}
Accretion-induced collapse (AIC) is a theoretically predicted final fate of white dwarfs (WDs). If a WD reaching the Chandrasekhar mass limit triggers electron-capture reactions at its center, the WD collapses to a neutron star (NS) \citep[e.g.,][]{nomoto1991}. This transformation from a WD to a NS is referred as AIC. AIC has been argued as a way to produce NSs in globular clusters \citep[e.g.,][]{bailyn1990} and some millisecond pulsars (e.g., \citealt{bhattacharya1991,tauris2013}). AIC can also be a site of the $r$-process nucleosynthesis \citep[e.g.,][]{qian2007} and ultrahigh-energy cosmic ray production \citep[e.g.,][]{piro2016}.

There are two major suggested evolutionary paths to cause AIC. The first path is through the accretion from non-degenerate stars onto O+Ne+Mg WDs [single-degenerate (SD) scenario, e.g., \citealt{nomoto1991}]. If the accretion rate (\Macc) is sufficiently high ($\Macc \sim 10^{-7}-10^{-5}~\Msun~\mathrm{yr^{-1}}$, e.g., \citealt{nomoto1991,nomoto2007,shen2007}), the O+Ne+Mg WDs can grow their mass close to the Chandrasekhar limit and the electron-capture reactions can be triggered at their center. The other path is through the merger of two WDs [double-degenerate (DD) scenario, e.g., \citealt{webbink1984,iben1984}]. WD mergers can lead to the formation of massive WDs depending on the mass ratio of the two WDs (e.g., \citealt{dan2014,shen2015}; \citealt{sato2016}). It has long been believed that the WD merger leads to the off-center carbon ignition \citep[e.g.,][]{nomoto1985,saio1985}. The carbon burning gradually propagates into the WD center, transforming a C+O WD into an O+Ne+Mg WD. If the WD formed by the merger is heavier than the Chandrasekhar mass limit, the O+Ne+Mg WD will eventually cause AIC when its center becomes dense enough because of the cooling \citep[e.g.,][]{yoon2005}. However, this classical view has also been recently questioned \citep[e.g.,][]{schwab2016,yoon2007}.

Understanding electromagnetic (EM) signatures of AIC and identifying them in transient surveys are important to confirm the existence of AIC. It is also important to know how often AIC actually occurs in the Universe to constrain the evolutionary path leading to AIC. Furthermore, AIC is a potential source of gravitational waves (GWs) \citep[e.g.,][]{abdikamalov2010} and the knowledge of their EM counterparts is essential in identifying them. Early studies by \citet{dessart2006,dessart2007} found that the radioactive \Ni, which is a major heating source of supernovae (SNe), is not much synthesized during AIC and no luminous optical transients may be accompanied by AIC. However, the subsequent studies by \citet{metzger2009,darbha2010} showed that AIC from rapidly rotating WDs may form a rotationally supported disk which makes \Ni-rich outflows and that AIC can be accompanied by faint optical transients evolving in a timescale of several days.

In this paper, we investigate EM properties of AIC in radio frequencies. Especially, since O+Ne+Mg WDs in SD systems cause AIC due to a large accretion, a large outflow from accreting WDs, companion stars, or their common envelopes is likely to exist as in Type~Ia SN progenitors from SD systems (e.g., \citealt{chomiuk2012} for a summary). The large outflow makes dense circumstellar media (CSM) and AIC occurs in the dense CSM. Then, the ejecta from AIC collide to the dense CSM and strong shock waves emitting radio can be formed. Radio emission due to the CSM interaction has been used to constrain the progenitor channel (SD or DD) of Type~Ia SNe. The lack of radio emission, and therefore dense CSM, in many Type~Ia SNe rules out most of the suggested SD models and the DD channel is favored in them \citep[e.g.,][]{chomiuk2016,chomiuk2012}, although some Type~Ia SNe show possible signatures of the SN-CSM interaction and they may be from the SD channel \citep[e.g.,][]{hamuy2003,foley2012,dilday2012}. However, in the case of AIC, the lack of the dense CSM in the DD channel does not prevent the appearance of strong radio emission. There is a chance to cause fast radio bursts (FRBs, e.g., \citealt{lorimer2007,thornton2013}) from the collapse itself as we discuss in this paper.

There are a few previous studies on radio transients from AIC. \citet{piro2013} suggest that spin-down of newly born magnetars by AIC makes a pulsar wind nebula (PWN) in the AIC ejecta and the interaction between the AIC ejecta and the PWN can result in radio emission. In this paper, we study radio emission due to the interaction between AIC ejecta and CSM created by their evolution towards AIC. \citet{metzger2015b} study the radio emission from AIC interacting with interstellar media, but not with CSM. No previous studies consider AIC as an FRB progenitor.

We first overview the predicted ejecta properties of AIC in Section~\ref{sec:aicprop}. Then, we discuss the radio emission expected from AIC in Section~\ref{sec:radiotran}. We discuss the rate and observations of the radio transients from AIC in Section~\ref{sec:rates} and conclude this paper in Section~\ref{sec:conclusions}.

\section{AIC ejecta properties}\label{sec:aicprop}
We first summarize properties of ejecta from AIC. The ejecta properties affect the radio emission from AIC especially when AIC occurs in the SD systems.

\citet{dessart2006} showed that AIC can result in explosions with the ejecta mass (\Mej) of $\sim 10^{-3}~\Msun$ and the explosion energy (\Eej) of $\sim 10^{49}~\mathrm{erg}$ with the neutrino-driven mechanism. Later, \citet{dessart2007} found that the ejecta mass and the explosion energy of AIC can be enhanced to $\Mej\sim 0.1~\Msun$ and $\Eej\sim 10^{51}~\mathrm{erg}$ if the AIC explosion is magnetically driven. In both cases, they find that the production of \Ni\ is negligible ($\sim 10^{-4}~\Msun$) and AIC is not optically bright.

However, subsequent studies by \citet{metzger2009,darbha2010} proposed a different mechanism in AIC to have \Ni-rich ejecta with $\Mej \sim 0.01~\Msun$. When AIC occurs in rapidly rotating WDs, an accretion disk supported by the centrifugal force can be formed. The disk initially becomes neutron-rich because of its neutrino emission. However, thanks to neutrinos from the proto-NS, the proton-to-neutron ratio in the disk becomes $\sim 1$ and the hot disk will be composed mostly of \Ni. The disk can eventually start to move outwards and become ejecta because of the viscous stress and the nuclear fusion energy provided by the $^4$He production. The typical ejecta velocity is estimated to be $\sim 0.1c$, where $c$ is the speed of light, and $\Eej$ is $\sim 10^{50}~\mathrm{erg}$. In this case, optical transients with the peak luminosity of $\sim 10^{41}~\mathrm{erg~s^{-1}}$ and the rise time of $1-10$~days can be accompanied by AIC.

Table~\ref{table:table} summarizes the three possible ejecta properties from AIC currently predicted. 

\begin{table}
\begin{center}
\caption{Predicted AIC ejecta properties.}
\begin{tabular}{ccccl}
\tableline\tableline
model & \Mej & \Eej & \Mni & reference \\
      & \Msun & $10^{51}~\mathrm{erg}$ & \Msun &  \\
\tableline
A & $\sim 10^{-3}$ & $\sim 0.01$ & $\sim 10^{-4}$ & \citet{dessart2006} \\
B & $\sim 0.1$ & $\sim 1$ & $\sim 10^{-4}$ & \citet{dessart2007} \\
C & $\sim 0.01$ & $\sim 0.1$ & $\sim 0.01$ & \citet{metzger2009} \\
\tableline
\end{tabular}
\label{table:table}
\end{center}
\end{table}

\begin{figure*}
 \begin{center}
  \includegraphics[width=\columnwidth]{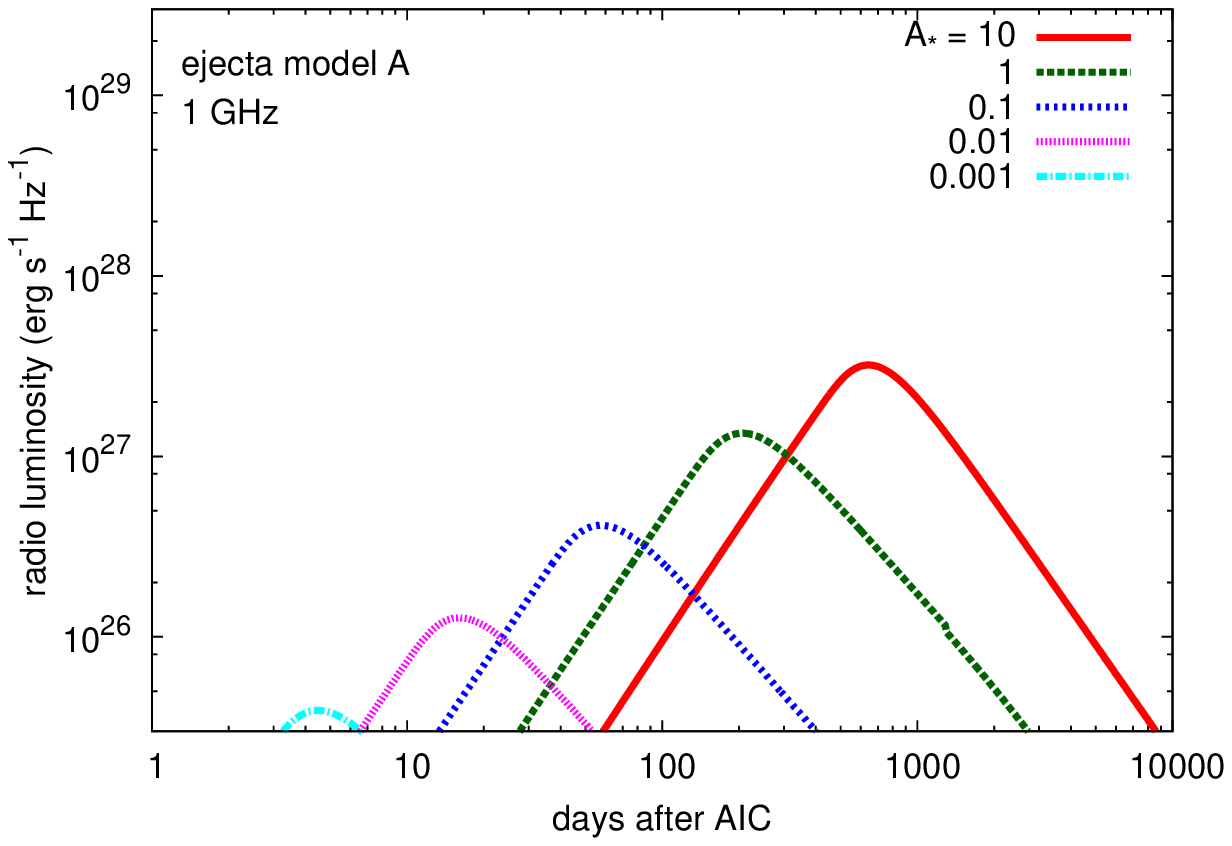}  
  \includegraphics[width=\columnwidth]{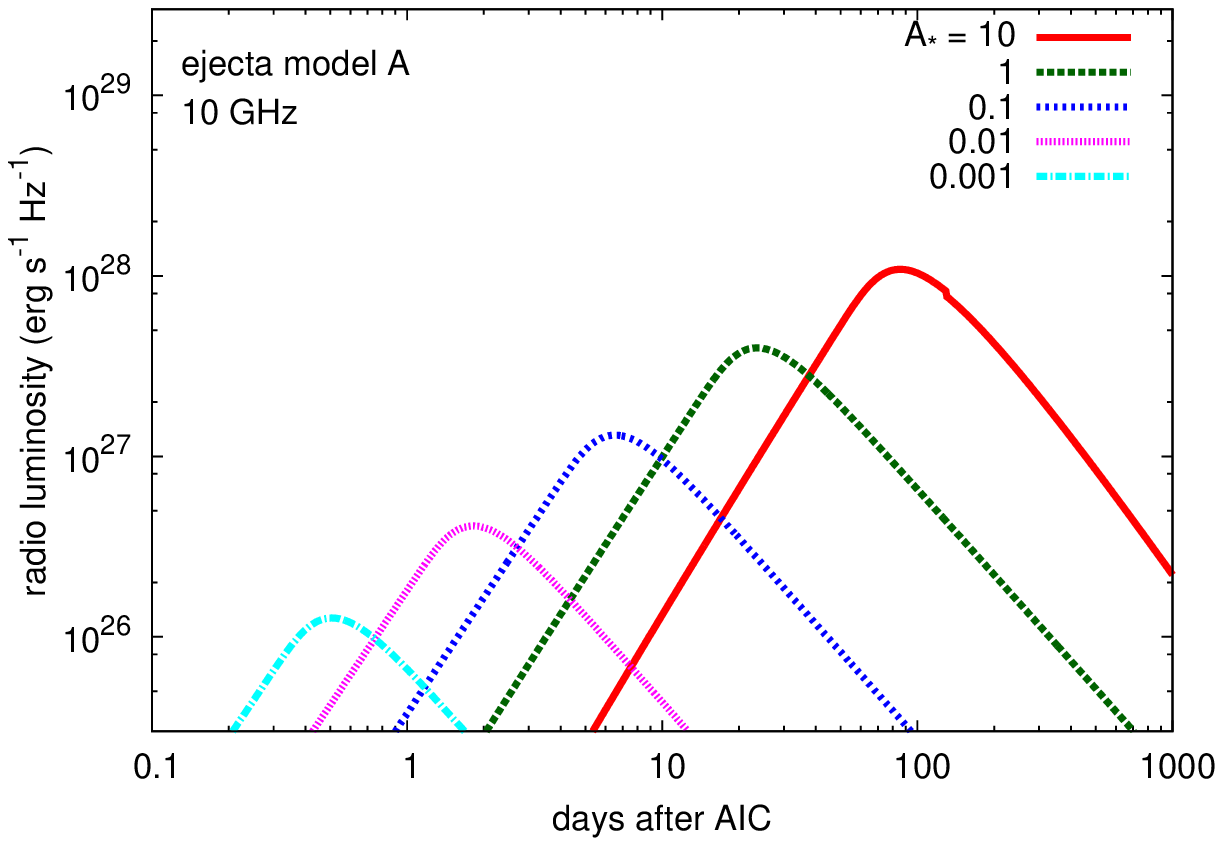} \\
  \includegraphics[width=\columnwidth]{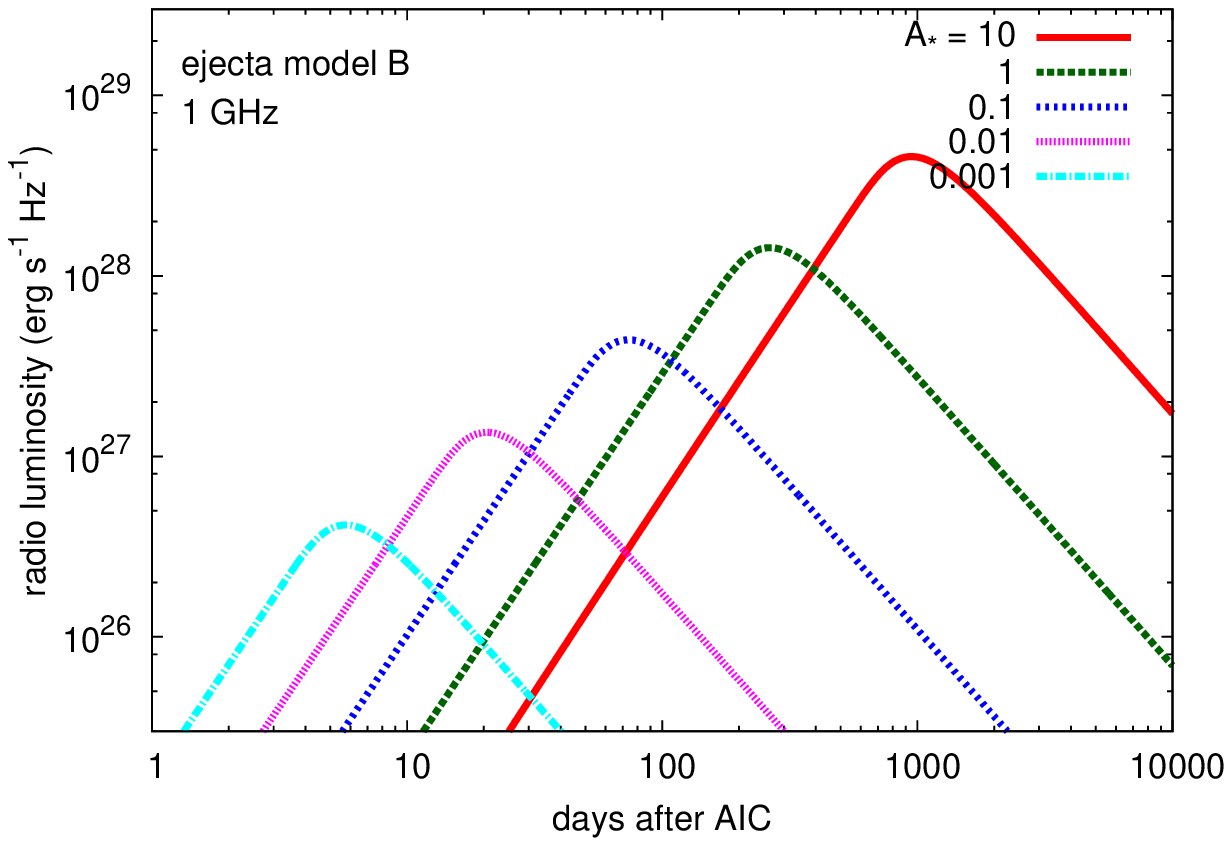}  
  \includegraphics[width=\columnwidth]{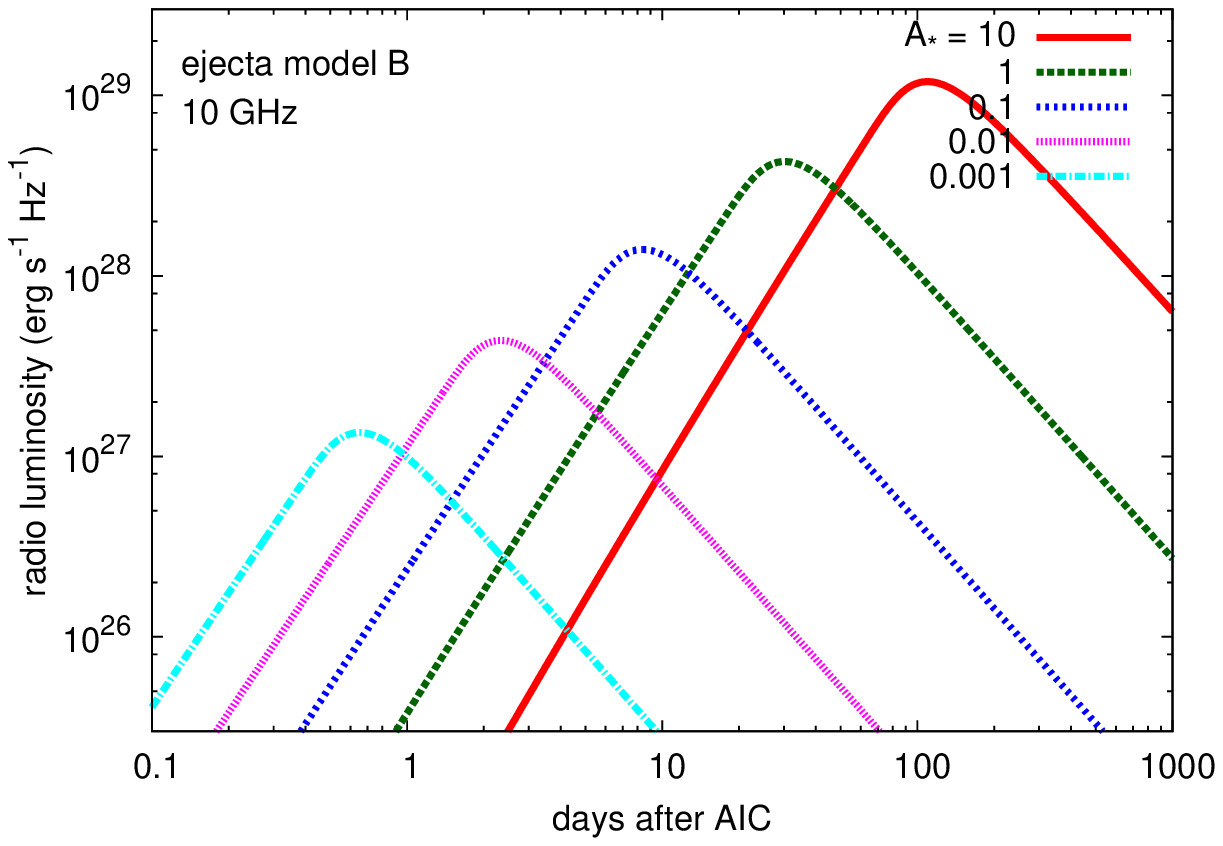} \\
  \includegraphics[width=\columnwidth]{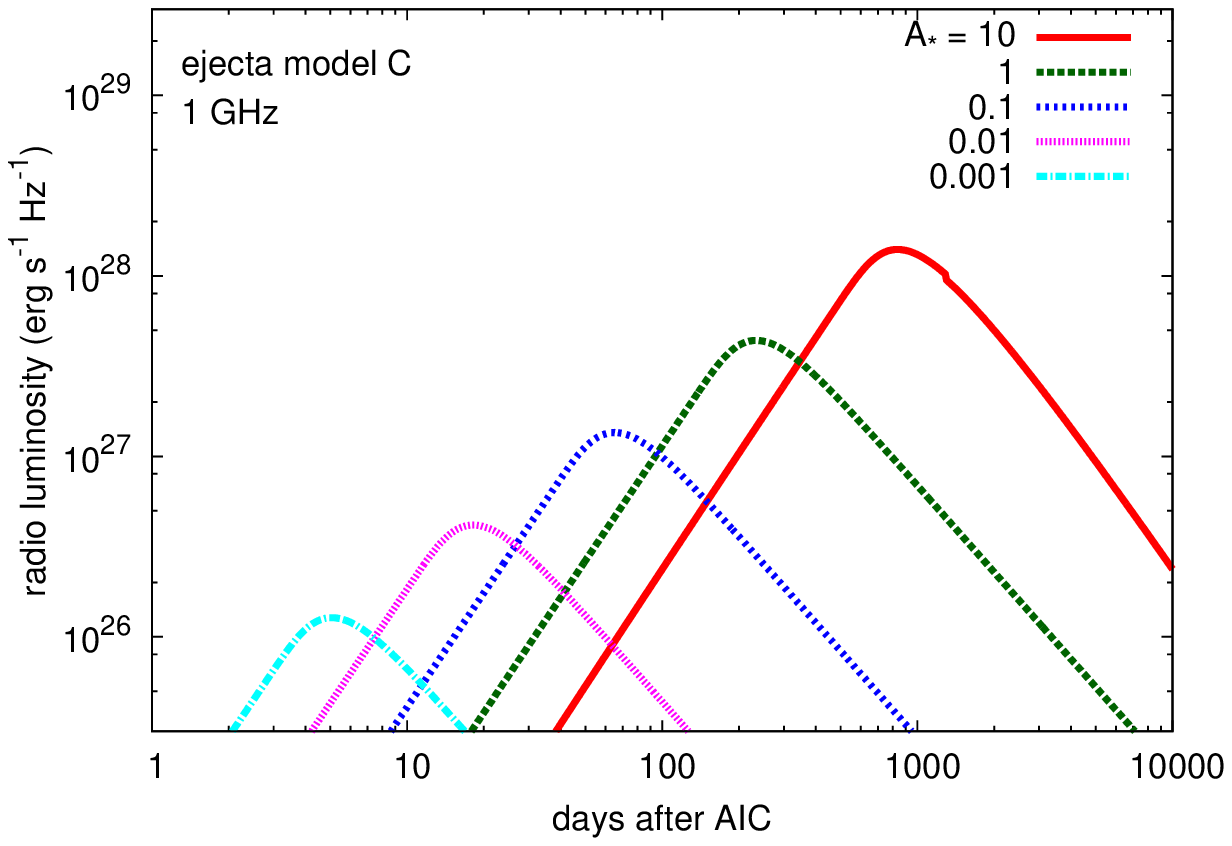}  
  \includegraphics[width=\columnwidth]{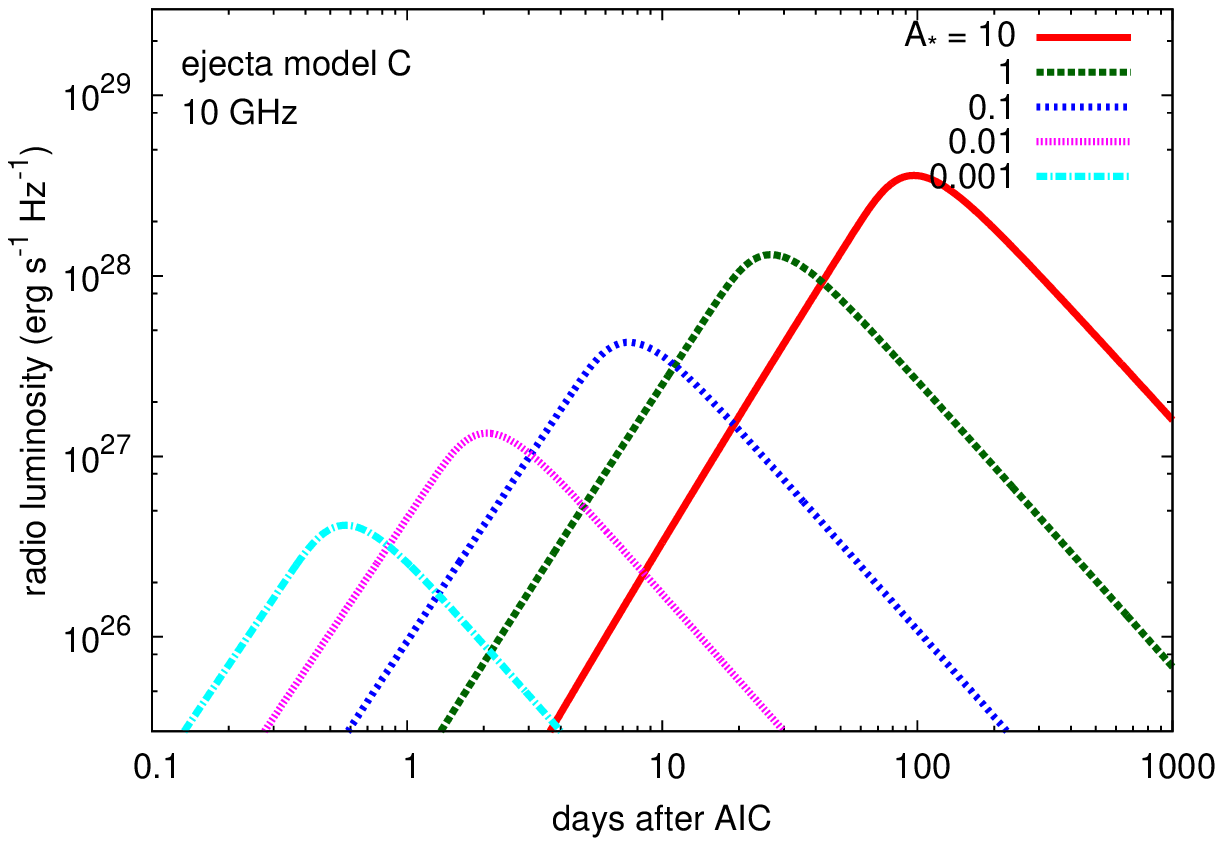} 
 \end{center}
\caption{
Radio LCs of AIC from SD systems at 1~GHz and 10~GHz with the different ejecta properties (Table~\ref{table:table}) and the different CSM densities ($A_*= 0.001-10$). Note the difference in the $x$ axis in the right and left panels.
}\label{fig:csmradio}
\end{figure*}

\section{Radio transients from AIC}\label{sec:radiotran}
\subsection{AIC from SD systems}
AIC in SD systems occurs if the accretion rate onto a WD is large enough to keep the surface burning steady. The required accretion rate is $\sim 10^{-7}-10^{-5}~\mathrm{\Msun~yr^{-1}}$ \citep[e.g.,][]{nomoto1991,nomoto2007,shen2007}. There are several SD evolutionary channels for WDs to achieve such a high accretion rate. The high accretion rates are often accompanied with high mass-loss rates, resulting in large CSM density. Thus, the ejecta from AIC collide with the dense CSM and a strong shock emerges. Electrons can be accelerated at the shock and synchrotron emission can be emerged from the accelerated relativistic electrons as found in SNe exploding within CSM \citep[e.g.,][]{fransson1998}. This synchrotron emission can be observed in radio frequencies.

Let \Mloss\ the mass-loss rate from a SD system leading to AIC. Then, the CSM density \rhocsm\ of the system is expressed as
\begin{equation}
\rhocsm (r) = \frac{\Mloss}{4\pi\vw}r^{-2},
\label{eq:csmdensity}
\end{equation}
where $r$ is the radius and \vw\ is the wind velocity. We assume that the ejecta expand homologously with the density structure $\rho_\ej\propto r^{-n}$ (outside) and $\rho_\ej\propto r^{-\delta}$ (inside). Then, the forward shock radius (\rsh) and velocity (\vsh) can be obtained analytically as a self-similar solution \citep{chevalier1982}. We adopt this analytic solution with $n=7$ and $\delta=1$. However, the analytic solution is no longer valid when the reverse shock starts to propagate into the inner flat density region. This starts when the swept-up CSM mass becomes comparable to \Mej, i.e., when $\Mej \sim \Mloss \rsh /\vw$ \citep[e.g.,][]{moriya2013}. After this moment, we assume that the shock expands following $\vsh = [2\Eej/(\Mej + \Mloss \rsh /\vw)]^{0.5}$. Although we change the formulation of $\vsh$ abruptly when $\Mej = \Mloss \rsh /\vw$ is satisfied for simplicity, \vsh\ is found to change smoothly. This transition only occurs in the highest CSM density models we show later in this section.

Given the shock radius $r_\sh$ and the shock velocity $v_\sh$, the synchrotron luminosity at a frequency $\nu$ from the shock $(L_\nu)$ can be estimated as (e.g., \citealt{bjornsson2004}; \citealt{fransson1998})
\begin{equation}
\nu L_\nu \simeq \pi \rsh^2\vsh n_\mathrm{rel,\gamma_\nu}
\left(\frac{\gamma_\nu}{\gamma_\mathrm{min}}\right)^{1-p}
\gamma_\nu m_e c^2
\left[
1+\frac{t_{\mathrm{sync},\nu}}{t}
\right]^{-1},
\label{eq:radiolum}
\end{equation}
where $n_\mathrm{rel,\gamma}$ is the number density of the relativistic electrons with the Lorentz factor $\gamma$, $\gamma_\nu=(2\pi m_ec\nu/eB)^{0.5}$ is the Lorentz factor of the relativistic electrons with the characteristic frequency $\nu$, $\gamma_\mathrm{min}\sim 1$ is the minimum Lorentz factor of the accelerated electrons, $m_e$ is the electron mass, $e$ is the electron charge, and $B$ is a magnetic field strength. We assume that the number density distribution of the relativistic electrons with the Lorentz factor $\gamma$ follows $dn_\mathrm{rel,\gamma}/d\gamma\propto \gamma^{-p}$ with $p=3$. The synchrotron cooling timescale at $\nu$ is $t_{\mathrm{sync},\nu}=6\pi m_e c/\sigma_T\gamma_\nu B^2$, where $\sigma_T$ is the Thomson scattering cross section. We also take the synchrotron self-absorption (SSA) at the shock into account \citep[e.g.,][]{chevalier1998}. The SSA optical depth is $\tau_\mathrm{SSA,\nu}=(\nu/\nu_\mathrm{SSA})^{-(p+4)/2}$ with the SSA frequency $\nu_\mathrm{SSA}\simeq 3\times 10^{5}(\rsh\epsilon_e/\epsilon_B)^{2/7}B^{9/7}$~Hz in cgs units. Here, $\epsilon_e$ is the fraction of the post-shock energy used for the electron acceleration and $\epsilon_B$ is the fraction converted to the magnetic field energy. We assume $\epsilon_e=\epsilon_B=0.1$ in the rest of this paper. We neglect the free-free absorption in the unshocked CSM as the CSM density is not high enough for it to be effective. 

The radio luminosity (Eq.~\ref{eq:radiolum}) is proportional to the CSM density. As the CSM density scales with $\Mloss/\vw$ (Eq.~\ref{eq:csmdensity}), the radio luminosity scales with $\Mloss/\vw$. Following the convention, we define $A_* \equiv (\Mloss/\vw)/(10^{-5}~\Msun~\mathrm{yr^{-1}}/1000~\mathrm{km~s^{-1}})$. The mass loss from the SD system determining $A_*$ depends on SD channels. We consider several possible outflows from the SD systems as follows (see, e.g., \citealt{chomiuk2012} for a summary).

One possible SD channel leading to a large accretion is the stable Roche-lobe overflow (RLOF) onto WDs. The RLOF can make nuclear burning stable at around the accretion rate required for AIC. Assuming that a small fraction ($\sim 1\%$) of the transferred mass is lost through the outer Lagrangian point (\citealt{chomiuk2012} and references therein), we expect $\Mloss\sim 10^{-9}-10^{-7}~\Msun~\mathrm{yr^{-1}}$ with $\vw\sim 100~\kmps$, or $A_*\sim 0.001-0.1$. However, the large accretion rate may result in the optically thick wind when the accretion rate is above $\sim 10^{-6}~\Msun~\mathrm{yr^{-1}}$ \citep{hachisu1999}. Then, $\Mloss\sim 10^{-6}-10^{-5}~\Msun~\mathrm{yr^{-1}}$ with $\vw\sim 1000~\kmps$ can be achieved. This outflow makes $A_*\sim 0.1-1$. Another possible SD channel is a symbiotic system in which the accreting mass is provided by a red giant \citep[e.g.,][]{seaquist1990}. The mass loss from the symbiotic system is dominated by the wind from the red giant. Assuming $\Mloss\sim 10^{-8}-10^{-6}~\Msun~\mathrm{yr^{-1}}$ with $\vw\sim 10~\kmps$, we expect $A_*\sim 0.1-10$. To summarize, $A_*\sim 0.001-10$ is expected in AIC from the SD model.

The radio LCs expected from the abovementioned SD systems are summarized in Figure~\ref{fig:csmradio}. The three different AIC ejecta in Table~\ref{table:table} are adopted. We show the radio LCs at two representative frequencies, 1~GHz and 10~GHz. The expected peak radio luminosity from AIC ranges $\sim 10^{26}-10^{29}~\mathrm{erg~s^{-1}~Hz^{-1}}$. The peak luminosities of the $A_*\sim 0.1$ models are comparable to those of stripped-envelope SNe, although stripped-envelope SNe typically have $A_*\sim 1$. This is because of higher velocities in AIC ejecta. While stripped-envelope SNe typically have $\Eej/\Mej\sim 1~\mathrm{foe}/\Msun$ where $1~\mathrm{foe}\equiv10^{51}~\mathrm{erg}$ \citep[e.g.,][]{lyman2016}, the AIC ejecta have $\Eej/\Mej\sim 10~\mathrm{foe}/\Msun$ (Table~\ref{table:table}). Therefore, their shock velocities $[\propto (\Eej/\Mej)^{0.5}]$ are higher, making the radio luminosities comparable to those of stripped-envelope SNe. The higher shock velocities also makes radio luminosities of AIC higher than those of Type~Ia SNe since $\Eej/\Mej$ is $\sim 1~\mathrm{foe}/\Msun$ in Type~Ia SNe as well \citep[e.g.,][]{mazzali2007,scalzo2014}. When $A_*\gtrsim 1$, the peak radio luminosities of AIC can be comparable to, or even higher than, those of Type~IIn SNe which are among the most radio luminous SNe \citep[e.g.,][]{perez-torres2015}. 

The rise times $t_\mathrm{rise,\nu}$ of the AIC radio LCs strongly depend on frequencies to observe, mainly because of the frequency dependence of SSA. The rise time follows $t_\mathrm{rise,\nu}\propto \nu^{-7(n-2)/(7n-12)}A_*^{(9n-22)/2(7n-12)}$ or $t_\mathrm{rise,\nu}\propto \nu^{-0.95} A_\star^{0.55}$ in our case. Roughly speaking, the rise time is inversely proportional to the frequency. Similarly, the peak radio luminosity is proportional to $\nu^{19/(7n-12)}A_*^{19(n-5)/2(7n-12)}$ or $\nu^{0.51}A_*^{0.51}$. Shortly, the radio LCs at longer frequencies have shorter rise times. The radio LCs at 1~GHz typically rise in $10-100$~days, while they typically rise within 10~days at 10~GHz. 

In the models with $A_\star = 10$, the deviation from the self-similar solution for \rsh\ and \vsh\ makes the LC decline after the peak steeper ($L_\nu \propto t^{-2}$). When the shock follows the self-similar solution, $L_\nu$ after the peak is proportional to $t^{-(n+1)/(n-2)}$ or $t^{-8/5}$ in our case.

Radio emission from AIC is also suggested to originate from the interaction between AIC ejecta and PWN \citep{piro2013}. It is important to distinguish the two different origins of the AIC radio emission observationally. \citet{piro2013} predict that the peak radio luminosity from the AIC-PWN interaction is $10^{28}-10^{29}~\mathrm{erg~s^{-1}~Hz^{-1}}$ at 1.4~GHz. The expected peak luminosity range for the AIC-CSM interaction is $10^{26}-10^{29}~\mathrm{erg~s^{-1}~Hz^{-1}}$ and the luminosity variation from the AIC-CSM interaction is larger. The radio LCs from the AIC-PWN interaction is likely to rise more quickly because the AIC ejecta are denser than the CSM and the free-free absorption can be stronger in early phases in the case of the AIC-PWN interaction. Finally, the radio luminosity from the interaction between AIC ejecta and the interstellar medium is expected to be fainter than that from the interaction between AIC ejecta and the CSM \citep{metzger2015b}.

\subsection{AIC from DD systems}\label{sec:dd}
Contrary to the case of AIC in SD systems, DD systems are not expected to have dense CSM. Therefore, the strong radio emission from the shock as in SD systems is not expected. However, we argue that AIC from DD systems has a possibility to cause FRBs.

The merger of two WDs can produce a rapidly rotating massive WD up to around 4~\Msun\ (e.g., \citealt{justham2011} and references therein). Therefore, AIC of massive WDs can form supramassive NSs which are heavier than $\sim 2.2~\Msun$ \citep[e.g.,][]{metzger2015} and require rapid rotation to support themselves. Such supramassive NSs can be a progenitor of FRBs if they are strongly magnetized \citep{falcke2014}. When their rotational energy goes below the minimum rotational energy required to support themselves, the strongly magnetized supramassive NSs collapse to BHs in a dynamical timescale \citep[e.g.,][]{shibata2000}. During the collapse, the strong magnetic field may make a burst in radio which can be observed as a FRB (so-called the ``blitzar'' model, \citealt{falcke2014}). It is usually assumed that the rotational energy is lost through the dipole radiation in the blitzar model, but strongly magnetized supramassive NSs created by AIC can immediately lose their rotational energy by the r-mode instability \citep{andersson1999}. Therefore, supramassive NSs from AIC can collapse to BHs shortly after AIC and AIC can be immediately accompanied by FRBs.

The supramassive NS model for FRBs is argued to be less favored because of the small number of magnetars observed in the Galaxy \citep{kulkarni2014}. However, in our scenario, the spin down of supramassive NSs is presumed to occur immediately after the AIC due to the r-mode instability, not due to the electromagnetic dipole radiation. Therefore, the magnetars resulting in FRBs can immediately collapse to BHs and the observed population of magnetars are not necessarily related to FRBs from supramassive NSs.

FRBs need to occur in a relatively clean environment because the radio signals can be absorbed by the surrounding environment if it is too dense. In the canonical AIC model leading to the NS formation, there are several suggested ways to make ejecta as discussed in the previous section. However, when AIC forms supramassive NSs, it is not clear whether there still remain ejecta as the BH formation can cease the ejecta formation. Therefore, FRBs from the DD system can occur in a clean environment and it is possible that they do not suffer from the strong absorption. Depending on the timescale of the BH transformation from the supramassive NSs, some ejecta may exist and may explain FRBs from relatively dense environment \citep{masui2015,kulkarni2015}.

FRBs are also suggested to occur during the merger of magnetized WDs \citep{kashiyama2013}. If FRBs from the NS merger and the subsequent AIC can successfully emerged, two FRBs may occur from the same progenitor system. Because the cooling timescale of the merged WD is $\sim 10^4$ years, the second FRB during AIC may occur $\sim 10^4$~years after the merger.

Finally, the discovery of a repeating FRB \citep{spitler2016,scholz2016} revealed that FRB models with cataclysmic phenomena cannot account for all FRBs. Therefore, our FRB model involving AIC can only explain a part of FRBs at most. However, the repeating FRB has some different features compared to the other FRBs and it is still possible that FRBs come from several progenitors including those involving cataclysmic phenomena.

\section{Event rates and prospects for future radio transient surveys}\label{sec:rates}
The predicted event rates of AIC are quite uncertain. AIC from SD systems is predicted to occur $\sim 10^{-4}-10^{-6}~\mathrm{yr^{-1}}$ in our Galaxy \citep[e.g.,][]{yungelson1998}. AIC from DD systems can be as much as $\sim 10\%$ of Type~Ia SNe, or $\sim 10^{-4}~\mathrm{yr^{-1}}$ in our Galaxy \citep{ruiter2009,yoon2007}. The AIC rate is also constrained by the amount of neutron-rich elements produced by AIC, and the overall AIC rate should be less than $\sim 10^{-4}~\mathrm{yr^{-1}}$ in our Galaxy \citep[e.g.,][]{fryer1999}.

A Galactic AIC rate $R_\mathrm{AIC} \equiv R_{-4} 10^{-4}~\mathrm{yr^{-1}}$ corresponds to a volumetric AIC rate of $10^3R_{-4}~\mathrm{Gpc^{-3}~yr^{-1}}$ \citep{metzger2009} if we assume that the AIC rate is proportional to the blue stellar luminosity \citep{phinney1991}. As we discussed, it is likely that $R_{-4}\lesssim 1$. Because the volumetric rate of FRBs is estimated to be $\sim 10^{4}~\mathrm{Gpc^{-3}~yr^{-1}}$ \citep[e.g.,][]{kulkarni2014}, the AIC may not account for all FRBs, although the FRB rate might be actually close to $\sim 10^{3}~\mathrm{Gpc^{-3}~yr^{-1}}$ \citep{totani2013}. However, the repeating FRB indicates that there can be several progenitors for FRBs \citep{spitler2016,scholz2016} and AIC may account for a fraction of FRBs. We also note that AIC from DD systems is naturally expected to occur in evolved galaxies as may be found for a FRB (\citealt{keane2016}, but see also, e.g., \citealt{williams2016,vendantham2016,akiyama2016} for the arguments against the host galaxy detection). Although it is argued that the blitzar model involving young magnetars is disfavored if a FRB host galaxy is old \citep{keane2016}, FRBs from the blitzar model can appear in old galaxies because a collapsing magnetar can be formed during the AIC after a WD merger as we proposed in Section~\ref{sec:dd}.

Many radio transient surveys have been performed to investigate the radio transient sky (see, e.g., \citealt{mooley2016} for a summary). AIC is an excellent target for radio transient surveys as it is particularly bright in radio (see also \citealt{metzger2015b} for discussion on the detection of radio transients from AIC). Assuming that AIC becomes brighter than $10^{28}~\mathrm{erg~s^{-1}~Hz^{-1}}$ (Figure~\ref{fig:csmradio}), we expect the AIC observational rate of $\sim10^{-4}R_{-4}~\mathrm{deg^{-2}~yr^{-1}}$ with the limiting brightness of 1~mJy at $\sim 1-10~\mathrm{GHz}$. Assuming a typical duration of 100~days, this rate roughly corresponds to the all-sky snap-shot detection number of $\sim 1\ R_{-4}~\mathrm{sky^{-1}}$. Given the low expected rates, it is likely that no previous radio transient surveys have detected AIC \citep[e.g.,][]{mooley2016}. This limit can be reached by a survey like Very Large Array Sky Survey with a proper cadence \citep{mooley2016}. If we can reach down to $1~\mu\mathrm{Jy}$, the expected AIC observational rate becomes $\sim 3R_{-4}~\mathrm{deg^{-2}~yr^{-1}}$ (or $\sim 3\times 10^4 R_{-4}~\mathrm{sky^{-1}}$) and this can be reached by planned radio transient surveys by Square Kilometer Array \citep[e.g.,][]{perez-torres2015}. Radio transient surveys synchronized with optical transient surveys are important to help identifying AIC, as AIC may be accompanied with faint or no optical transients. Such radio transient surveys are important to directly confirm the existence of AIC, to constrain the AIC rate, and to unveil the major evolutionary path to AIC.

It is also important for radio transient surveys to perform coordinated observations with GW observatories to look for possible radio counterparts of GW sources, as AIC is a possible GW emitter \citep[e.g.,][]{abdikamalov2010}. Because the rise times of the accompanied radio transients are shorter in higher frequencies (Figure~\ref{fig:csmradio}), the radio follow-up observations in higher frequencies are easier to associate the detected transients to the GW sources. Although the poor source localization ability of the current GW observatories makes it hard to perform follow-up EM observations \citep[e.g.,][]{abbott2016}, the localization will be improved with the upcoming additional GW observatories.

\section{Conclusions}\label{sec:conclusions}
We have studied AIC as a progenitor of radio transients. When AIC occurs due to accretion onto an O+Ne+Mg WD from a companion star (the SD model), a dense CSM is likely to exist around the progenitor because of the large accretion rates required for AIC. Therefore, the ejecta from AIC collide with the dense CSM and a strong shock emerges. The shock can emit bright synchrotron radiation which is observable in radio frequencies. We found that the peak radio luminosity from such progenitor systems can be $\sim 10^{26}-10^{29}~\mathrm{erg~s^{-1}~Hz^{-1}}$ at $\sim 1-10~\mathrm{GHz}$, depending on the SD channel, and the typical rise times are $\sim 10-1000$~days in 1~GHz and $\sim 1-100$~days in 10~GHz. Because AIC is likely accompanied by faint optical transients whose peak luminosities are $\sim 10^{41}~\mathrm{erg~s^{-1}}$, AIC is observed as radio-bright optically-faint transients. Although we focused on radio transients in this Letter, the strong interaction between dense CSM and AIC ejecta is also likely to result in strong X-ray emission.

AIC can also occur after the merger of WDs if the merger results in the formation of a WD heavier than the Chandrasekahr mass limit (the DD model). Although there do not exist dense CSM in AIC from the DD model, AIC of the massive WD may result in a FRB. If a strongly magnetized supramassive NS is formed by AIC, the NS may quickly lose its rotational energy required to support itself because of the r-mode instability and it may immediately collapse into a BH. Then, the collapsing strongly magnetized NS may cause a FRB through the ``blitzar'' mechanism \citep{falcke2014}. Because the merger of strongly magnetized WDs is also suggested to make a FRB \citep{kashiyama2013}, two FRBs separated by $\sim 10^4$~years may occur in the same progenitor.

The rates of AIC from both the SD and DD systems are likely less than $\sim 10^{3}~\mathrm{Gyr^{-3}~yr^{-1}}$. The observed FRB rates ($\sim 10^4~\mathrm{Gyr^{-3}~yr^{-1}}$, e.g., \citealt{kulkarni2014}) are higher than the expected AIC rate and only a fraction of FRBs can be from AIC. The radio transient surveys with the limiting radio luminosity of 1~mJy are expected to detect AIC from the SD model with the rate of $\sim 10^{-4}~\mathrm{deg^{-2}~yr^{-1}}$ $(\sim 1~R_{-4}~\mathrm{sky^{-1}})$ or less. Radio transient surveys like Very Large Array Sky Survey have possibilities to detect radio transients with such a rate \citep[e.g.,][]{mooley2016}. If radio transient surveys with the limiting luminosity of $1~\mu\mathrm{Jy}$ are performed as planned by Square Kilometer Array \citep[e.g.,][]{perez-torres2015}, we expect to detect AIC with the rate of $\sim 3~\mathrm{deg^{-2}~yr^{-1}}$ $(\sim 3\times 10^{4} R_{-4}~\mathrm{sky^{-1}})$ or less. These radio transient surveys enable us to directly confirm the existence of AIC, to constrain the AIC rate, and to know the major progenitor path leading to AIC.

AIC is also suggested to be GW sources \citep[e.g.,][]{abdikamalov2010}. It is essential to understand the EM counterparts of GWs to identify their origins. We have shown that GWs from AIC can be accompanied by radio-bright optically-faint (possibly X-ray-bright) transients. Because the rise times of the AIC radio LCs are much shorter in the longer frequencies, the follow-up radio observations in longer frequencies can be more beneficial to identify the EM counterparts of GW sources in radio frequencies.

\acknowledgments{
I would like to thank the referee for the constructive comments that improved this work significantly. This research is supported by the Grant-in-Aid for Research Activity Start-up of the Japan Society for the Promotion of Science (16H07413).
}

\bibliographystyle{yahapj}

\end{document}